\documentclass[jap,aip,reprint,longbibliography,amsmath,amssymb,twocolumn,superscriptaddress]{revtex4-1}
\usepackage{graphicx}
\usepackage{bm}
\usepackage{amsmath}
\usepackage{indentfirst}
\usepackage{times}
\usepackage{verbatim}
\usepackage{float}
\usepackage{color}
\usepackage{xcolor}
\usepackage{tabularx}

\begin{document}

\title{Identification of novel organic polar materials: A study combining importance sampling with machine learning}

\author{Ayana Ghosh}
\email{ghosha@ornl.gov}
\affiliation{ Center for Nanophase Materials Sciences and Computational Sciences and Engineering Division, Oak Ridge National Laboratory, Oak Ridge, Tennessee 37831, USA}
\affiliation{Department of Materials Science $\&$ Engineering and Institute of Materials Science, University of Connecticut, Storrs, Connecticut 06269, USA}
\author{Dennis P.\ Trujillo}
\affiliation{Department of Materials Science $\&$ Engineering and Institute of Materials Science, University of Connecticut, Storrs, Connecticut 06269, USA}
\author{Subhashis Hazarika}
\affiliation{Palo Alto Research Center, Palo Alto, California 94304, USA}
\author{Elizabeth Schiesser}
\affiliation{\hbox{Department of Biomedical Engineering, University of Connecticut, Storrs, Connecticut 06269, USA}}
\author{M. J. Swamynathan}
\affiliation{\hbox{Department of Physics and Nanotechnology, Faculty of Engineering and Technology, SRM Institute of Science and Technology, Kattankulathur - 603 203, Tamil Nadu, India}}
\author{Saurabh Ghosh}
\affiliation{\hbox{Department of Physics and Nanotechnology, Faculty of Engineering and Technology, SRM Institute of Science and Technology, Kattankulathur - 603 203, Tamil Nadu, India}}
\author{Jian-Xin Zhu}
\affiliation{\hbox{Theoretical Division, Los Alamos National Laboratory, Los Alamos, New Mexico 87545, USA}}
\affiliation{\hbox{Center for Integrated Nanotechnologies, Los Alamos National Laboratory, Los Alamos, New Mexico 87545, USA}}
\author{Serge Nakhmanson}
\email{serge.nakhmanson@uconn.edu}
\affiliation{Department of Materials Science $\&$ Engineering and Institute of Materials Science, University of Connecticut, Storrs, Connecticut 06269, USA}
\affiliation{Department of Physics, University of Connecticut, Storrs, Connecticut 06269, USA}

\date{\today}

\begin{abstract}
Recent advances in the synthesis of polar molecular materials have produced practical alternatives to ferroelectric ceramics, opening up exciting new avenues for 
their incorporation into modern electronic devices. 
However, in order to realize the full potential of polar polymer and molecular crystals for modern technological applications, it is paramount to 
assemble and evaluate all the available data for such compounds, identifying descriptors that could be associated with an emergence of ferroelectricity. 
In this work, we utilized data-driven approaches to judiciously shortlist candidate materials from a wide chemical space that could possess ferroelectric functionalities. 
An importance-sampling based method was utilized to address the challenge of having a limited amount of available data on 
already known organic ferroelectrics.
Sets of molecular- and crystal-level descriptors were combined with a Random Forest Regression algorithm in order to 
predict spontaneous polarization of the shortlisted compounds.
First-principles simulations were performed to further validate the predictions obtained from the machine learning model.
\end{abstract}

%\pacs{}

\maketitle

\section{introduction}\label{intro}
\begin{figure}
\centering
\includegraphics[width=1.0\columnwidth]{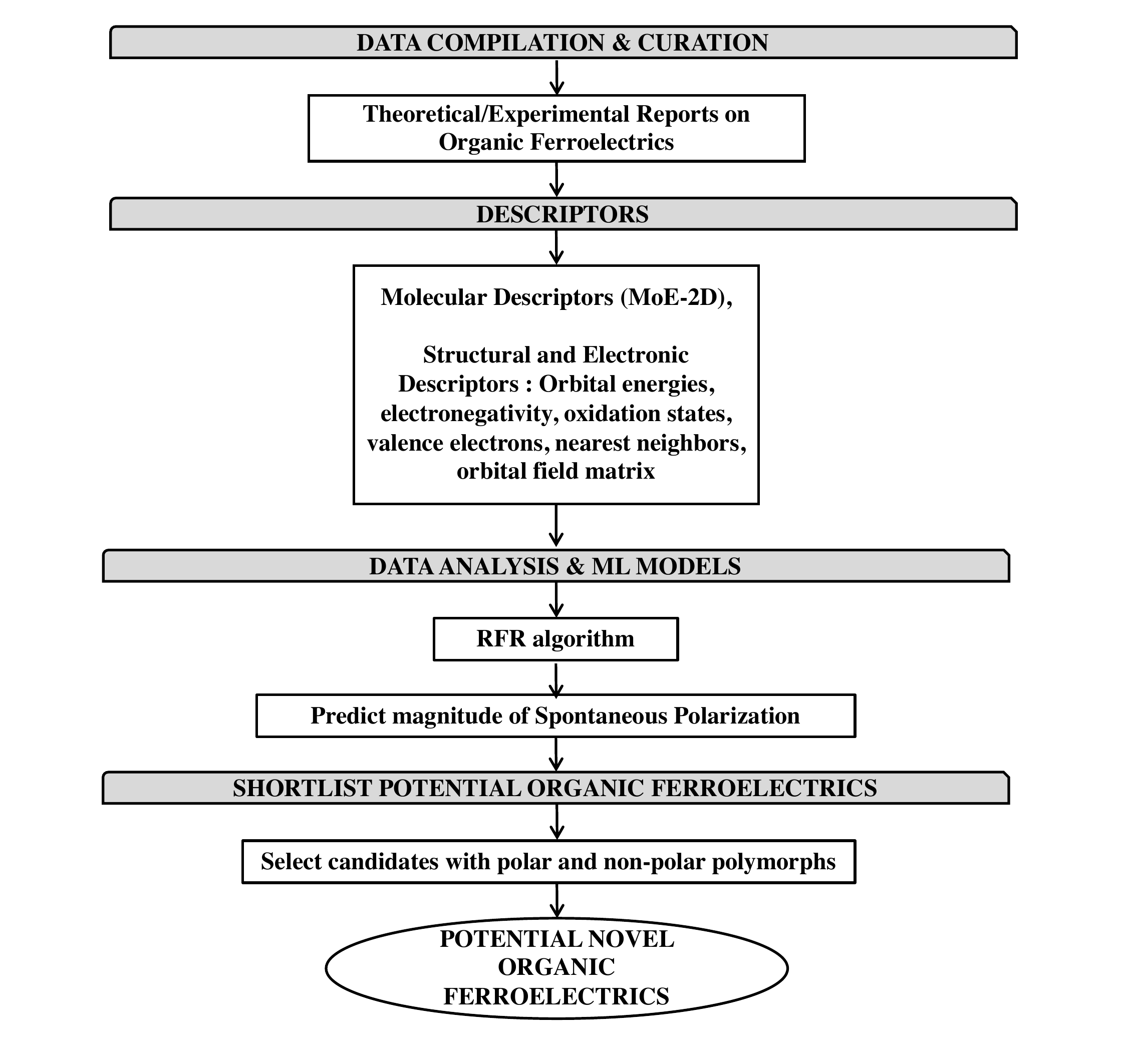}
\caption{
A flowchart of main stages involved in development of the selection process for
identifying molecular- and polymer-based materials with ferroelectric propensity. 
Primary stages are shown as grey rectangles, while any necessary secondary stages are represented by white rectangles.
All methodological details involved in implementation of each stage are presented in Sec.~\ref{methods}.
}
\label{fig:wf_orgferro}
\end{figure}

Ferroelectric materials are widely employed in a variety of technological applications, including acoustic, nonlinear-optic
and electromechanical devices --- i.e., sensors, actuators and transducers --- as well as in pyroelectric arrays, nonvolatile memories, and
high-$k$ dielectric components for microelectronics. 
Perovskite ABO$_3$-type metal-oxide ceramics, such as BaTiO$_3$, PbTiO$_3$, KNbO$_3$, or related solid solutions, such as
PbZr$_{1-x}$Ti$_x$O$_3$ (PZT), usually exhibit the strongest spontaneous polarization, as well as piezoelectric, 
and pyroelectric response.
However, the usefulness of oxide ceramics is limited by their substantial weight, brittleness, toxicity (e.g., in lead-containing 
compounds) and considerable costs of material synthesis and device manufacturing.\cite{choi2004enhancement,lee2005strong}
On the other hand, molecular- or polymer-based ferroelectrics, while displaying more modest polar properties, 
are usually lightweight, flexible and environmentally friendly.\cite{lovinger83,furukawa89,kepler92,nalwa95,eberle96,samara01}
A wide range of practical and inexpensive prescriptions is available for their growth and processing as either bulk materials or
nanostructures.
In order to fully realize the potential of polar polymer- and molecular-based materials for modern technological applications, 
it is paramount to acquire detailed molecular-level understanding of mechanisms governing the emergence of ferroelectricity in them.
It is noteworthy that some of the most well known ferroelectric compounds,\cite{horiuchi2008organic} e.g.,
Rochelle salt, in which ferroelectricity was originally discovered, or NaNO$_2$
that exhibits incommensurate structural distortions,\cite{heine1984origin} can be classified as molecular crystals.
Compared to the transition-metal oxide counterparts, the origins of ferroelectric behavior in such materials are more 
diverse and not as exhaustively investigated.\cite{louis2018polarization} 
Therefore, reliable paths to enhancement and control of their useful properties still largely remain to be charted out.
Despite the existence of various physical mechanisms that
could potentially induce ferroelectricity in polymer and molecular-based materials
--- in addition to the displacive mechanism which dominates in ceramics --- it continues to be
rare, a `one-off' or spurious effect.
Lessons learned by some of the authors when studying DIPA-X \cite{louis2018polarization} and PVDF-based \cite{ghosh2018first} 
families of electroactive compounds provided some insight into why this may be so. 
Even though individual molecular units may be highly polar, when they are assembled into a periodic crystal structure, 
there is no guarantee that it will be ferroelectric: [A] Unit arrangements promoting cancellation of large individual dipole moments
are likely to form, resulting only in weak or complete absence of polarization; [B] Built-in energy barriers for dipole reorientation may
be too high to allow `easy' polarization switching among multiple possible directions.
In the latter case, the material may formally be ferroelectric, but the associated enormous coercive fields make
applications impractical.
Although specific pieces of knowledge useful for successful cultivation of ferroelectric properties in
polymer and molecular-based systems may remain elusive --- unquestionably, because of the structural richness of the 
involved material families and great complexity of the underlying physical mechanisms ---
this problem is highly amenable to probing with generic machine learning (ML) and data mining approaches.\cite{Sattari2021}
The main idea is to utilize the available information for the already known ferroelectric molecular-  
and polymer-based compounds to uncover any general commonalities in and connections among their
structure, chemical makeup, properties and performance, identifying a small number of simple
descriptors that `point' to ferroelectric behavior.
These descriptors can then be used to predict the likelihood of such behavior in other molecular- and polymer-based compounds, 
whether fictitious or already synthesized.
Engaging in such a study involves a number of difficulties, with the most obvious being:
(\emph{a}) Scarcity of the available data on the already known molecular and polymeric ferroelectrics, as
there are just not too many of them --- less than a hundred, which stipulates the
types of ML algorithms that could used; 
(\emph{b}) Lack of any standards on how such data is organized and curated,
i.e., individual research groups report whatever they feel is important in each particular case, rather than 
a prescribed set of descriptors, plus different groups have dissimilar quality standards for what is
considered `reliable data.'
On the other hand, for compound crystal systems that could be readily separated into
individual molecule-based structural units, preexisting libraries of numerous
chemical descriptors could be readily applied for characterization of these units.
Naturally, such chemical descriptors acting on the molecular level can be combined with global,
crystal-level descriptors.
Guiding principles for the selection of prospective ferroelectric materials:
For the purposes of this, or a similar (more detailed) investigation, the following natural connections between the material 
structure / properties and the desired `ferroelectric propensity' could be used for shortlisting the
most attractive target compounds from a larger pool of possible candidates.
(\emph{i}) Existence of multiple polymorphs for the same compound, some of which are non-centrosymmetric 
and polar, while others are centrosymmetric and non-polar. 
(\emph{ii}) Energetic closeness of the polar and aristotype non-polar phases,
meaning that these phases cannot be drastically structurally different. 
(\emph{iii}) Existence of well-behaved transformation paths connecting different variants of the polar 
phases, indicating feasibility of ferroelectric switching among them.
While criterion (\emph{i}) is based purely on symmetry considerations and thus is easy to include into a data-driven
workflow, criteria (\emph{ii}) and (\emph{iii}) may require deployment of extra and possibly substantial computational efforts, 
e.g., to evaluate relative energy differences between the phases, as well as probe possible transformation paths connecting them.
In this study, we aimed to avoid, whenever possible, doing any heavy atomistic-level calculations of this sort.
Although, clearly, for each prospective ferroelectric identified, these (and other) calculations would be
necessary to characterize its behavior with confidence, here our goal was to spin out
multiple (but not too many!) reasonable guesses for such prospective compounds and provide rough, but
computationally cheap estimates of their likely properties.
In what follows, we demonstrate how we have addressed some of the challenges stated above, showcasing an original 
data-driven and ML approach that can judiciously choose target materials likely to exhibit ferroelectricity from a wide chemical space 
of different molecular- and polymer-based candidate compounds. 
In conjunction, we present a number of crystal- and molecular-level descriptors that were recognized as important for predicting ferroelectric behavior.
In addition, we highlight the utilization of traditional ML algorithms for constructing a model that can estimate the magnitude of spontaneous 
polarization in the identified target materials without performing computationally expensive atomistic-level calculations. 
The flowchart showing connections between the main developmental stages of this study is presented in Fig.~\ref{fig:wf_orgferro}.

\section{Methods}\label{methods}
\subsection{Data Compilation and Curation}
As a starting point of this study, a dataset, called in what follows `Dataset I,' was compiled and curated, 
containing data records for most already known molecular- and polymer-based ferroelectrics  
reported in the literature.\cite{horiuchi2008organic,lines2001principles,ye2018metal}
Table~I in the Supplementary Material lists stoichiometric formulas and SMILEs (simplified molecular-input line-entry system)
strings for molecular units of all of the collected entries, with the latter 
providing more detailed description for the compound structure and chemical makeup.
Other properties collected by curating the literature were lattice parameters for unit cells, molecular weight (g/mol), 
melting temperature ($^\circ$C), density (g/cm$^3$),
polarization values (C/m$^2$), transition temperature ($^\circ$C) and space groups.
We note that for some molecular systems present in Dataset I the complete information could not be obtained. 
Therefore, the initial descriptor space for these compounds was built using only SMILEs representations, while 
assuming cubic unit cells.
Ideally, descriptions of the structure of these molecular systems should be clarified 
(e.g., by using first-principles theory computations for structural optimization), but
such efforts are beyond the scope of this work.
At the time of this writing, Dataset I contains information for a total of 76 molecular systems, 
with reported origins for the emergence of ferroelectricity including electronegativity differences between atoms, 
secondary interactions such as hydrogen bonding, as well as lattice-distortion driven structural instabilities.
The chemical nature and structure of the dataset entries are extremely diverse and include 
molecular and polymer crystals, oligomers, co-crystals, clathrates, charge-transfer complexes and
derivatives of metal-organic frameworks.
Only systems that have been reported by multiple research groups and could be verified by comparison with 
common databases --- in particular, Cambridge Structural Database (CSD)\cite{groom2016cambridge} ---
were included in Dataset I.
CSD utilizes the R-factor metric to determine the quality of structural data, with the average value of considerable 
R-factor for small organic molecules (like ones used in Dataset I and for shortlisting potential organic ferroelectrics) 
being $<$ 0.05.

\subsection{Descriptors}
For each of the Dataset I compounds, we have computed sets of molecular- and crystal-level descriptors.
For the former, standard computational Molecular Operating Environment (MOE)\cite{clark20062d,Moepackage} 2D chemical descriptors 
were used, while descriptors relying on a three dimensional shape of the investigated molecule were excluded 
to avoid adding any potential ambiguities to the descriptor space.
Crystal-level descriptors, as implemented in a Python library,\cite{matminer,ricci2020gapped,ward2016general} included atomic orbital
information, i.e., energies of highest occupied molecular orbital (HOMO) and lowest unoccupied molecular orbital (LUMO), 
orbital energies, types of and distances between atomic sites, fractions of nearest neighbors 
for each atomic and bond type.
Statistical data on electronegativity differences between anions and cations, 
oxidation states of all atoms, valence orbital attributes, such as mean number of electrons in each shell, and orbital field
matrix descriptions of the chemical environment of each atom in the unit cell, based on the group numbers, row numbers,
distances between coordinating atoms, and Voronoi polyhedra weights were also included.

\subsection{ML Models for Predicting Polarization}
A family of ML models was constructed utilizing the computed sets of molecular- and crystal-level descriptors.
Initial models included all the descriptors, while consecutive ones were built on only 10 most important descriptors (combined from both sets) 
as identified by the ML algorithm.
The models are constructed based on the best set of hyperparameters after performing full grid-search optimization.
The main target property to be predicted was the magnitude of spontaneous polarization in the compound / system of interest.
The Random Forest Regression (RFR) algorithm was employed as implemented in scikit-learn \cite{scikitlearn} to construct the models.
In general, regression-based algorithms were preferred over classification-based ones, as the endpoint of interest --- such as, e.g., 
magnitude of the spontaneous polarization --- is continuous in nature.
The choice of RFR, in comparison to other regression-based algorithms, was motivated by its performance 
in the previous studies\cite{ghosh2020machine, ghosh2019assessment} when applied to datasets of small size, as well as its 
capability to rank utilized descriptors by importance.
This feature of the RFR algorithm proved to be essential in this investigation, allowing us to identify the likely physical underpinnings 
that enable ferroelectricity in molecular and polymeric systems.

\subsection{Descriptor Ranking for Polarization Prediction}

An analysis of the relative importance of each descriptor (belonging to either molecular- or crystal-level descriptor set) 
was performed utilizing the RFR algorithm feature importance ranking to shortlist the descriptors useful for predicting polarization. 
The feature importance score produced by RFR represents the decrease in the weighted impurity 
(variations for regression type of algorithm) over all trees for every feature.
As determined, top ten ranked molecular- and crystal-level descriptors provide some physical insight into the structural 
and chemical properties that may influence polar behavior. 

\begingroup
\begin{table*}
\centering 
\caption{Top ten crystal-level descriptors important for predicting polarization as determined by the RFR
algorithm feature importance ranking of the descriptor set.
Descriptors are ordered in descending order of importance.
}\label{table1crd}
\begin{tabular}{ll}
\hline 
Descriptor	& Extended Definition  \\
\hline
MagpieData mean NdUnfilled & Mean number of unfilled $d$ valence orbitals among elements \\
MagpieData minimum GSbandgap &  Minimum value of the DFT bandgap of elemental solid among elements in composition  \\
vpa & Volume per atom of the crystal structure \\
MagpieData range Row & Range of the row on periodic table which each element in the composition belongs to  \\
MagpieData avg dev NUnfilled & Average deviation of unfilled valence orbitals among elements  \\
MagpieData avg dev NpUnfilled & Average deviation of number of unfilled $p$ orbitals among elements in composition  \\
MagpieData mean Electronegativity & Mean value of electronegativity among elements in composition \\
MagpieData avg dev NpValence & Average deviation of unfilled $p$ valence electrons among elements  \\
MagpieData avg dev NdValence & Average deviation of filled $d$ orbitals among elements in composition  \\
\hline
\end{tabular}
\end{table*}
\endgroup

For the crystal-level descriptors, it was observed that descriptors representing valence electrons and
elemental electronegativity in composition contributed greatly to predicting the polarization of 
molecular compounds (see Table~\ref{table1crd}).
This is expected since differences in electronegativity between neighbor atoms give rise to local dipole moments 
which then sum up to produce net polarization in the crystal environment.
These descriptors can be further related to the presence of nitrogen, hydrogen and oxygen bonds with metals, such as rhenium and bromine, 
and, therefore, the propensity of these particular elements to exhibit a variety of highly electropositive oxidation states. 
Finally, descriptors representing the crystal volume, average number of unfilled $p$ and $d$ orbitals, as well as deviations among them
in a molecular crystal were shown to be relatively important to predicting polarization. 

\begingroup
\begin{table*}
\centering
\caption{Top ten molecular-level descriptors important for predicting the polarization as determined by the RFR
algorithm feature importance ranking of the descriptor set.
Descriptors are ordered in descending order of importance.
}\label{table2mold}
\begin{tabular}{ll}
\hline 
Descriptor & Extended Definition \\
\hline
ATS1m &	Broto-Moreau autocorrelation weighted by mass  \\
GATS1v &	Geary autocorrelation weighted by van der Waals volumes \\
ATSC1p &	 Centered Broto-Moreau autocorrelation weighted by probabilities \\
maxsssCH &	Max atom type electronic state -CH- \\
ETA\_dPsi\_B & Measure of hydrogen bonding propensity  \\
ATSC2p &	 Centered Broto-Moreau autocorrelation weighted by probabilities 	\\
maxsI &	Max atom type electronic state for I \\
SpMax3\_Bhm & Largest absolute eigenvalue of Burden modified matrix wighted by relative mass \\
AATSC2m	 & Average centered Broto-Moreau autocorrelation weighted by mass  \\
AATS1m & Average Broto-Moreau autocorrelation weighted by mass  \\
\hline
\end{tabular}
\end{table*}
\endgroup

Relative importance of the molecular-level 2D descriptors was also determined in the same fashion. 
As shown in Table~\ref{table2mold}, descriptors representing topological distribution of mass, van der Waals volumes, 
and probabilities (associated with Broto-Moreau and Geary autocorrelations\cite{arthur2016quantum} of these parameters)  
were identified as important for predicting polarization.
Other important features identified by this analysis included electronic states related
to C-H and I bonding, and molecular propensity to form hydrogen bonds. 
This indicates that the presence of excess H atoms, as well as the presence of iodine (which is highly electronegative), 
contributes to the polar strength of molecular units comprising the crystal. 
We note that the molecular-level set descriptors were not considered in the final model for predicting polarization,
since they were not found to be highly important for the target property prediction.

\subsection{Data Analysis and Improvement of the ML Models}\label{MLM_imp}
\begin{figure}
\centering
\includegraphics[width=\columnwidth]{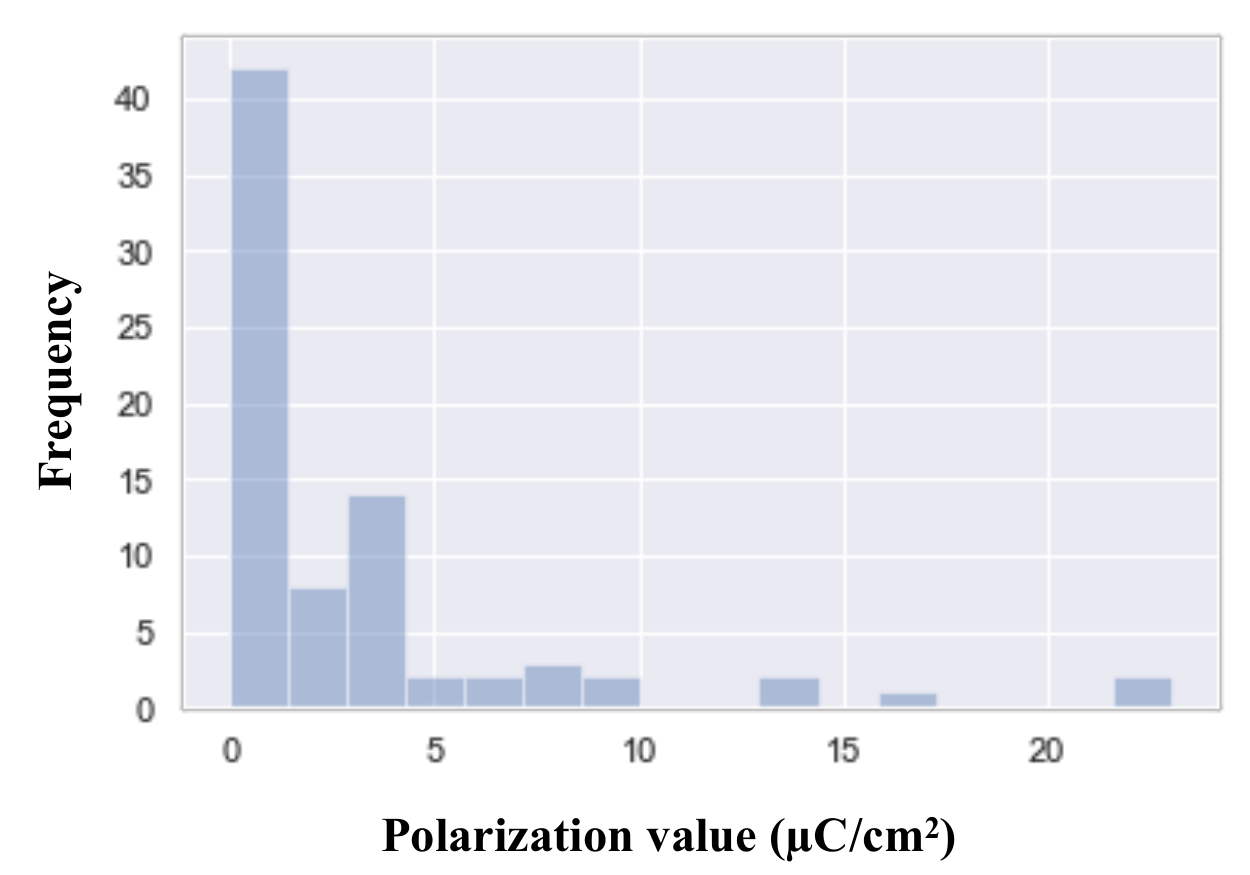}
\caption{
Histogram partitioning of compounds in Dataset I by the magnitude of their spontaneous polarization.}
\label{fig:fig2_org_ferro}
\end{figure}

A histogram plot of reported spontaneous polarizations exhibited by the compounds comprising Dataset I is 
shown in Fig.~\ref{fig:fig2_org_ferro}.
From this data, it is evident that in addition to being limited in size, Dataset I 
is strongly skewed or biased towards systems with rather low polarization magnitudes. 
That is not entirely unexpected, considering that molecular- or polymer-based ferroelectric compounds usually 
have smaller polarization, compared to their ceramic counterparts.
In this situation, an infusion of a substantial number of additional data entries representing systems possessing high 
spontaneous polarization is required to remove the bias in the data. 
In general, training predictive ML models on well-balanced data produces better prediction accuracy,
compared to using data that is unbalanced.\cite{imbalancedDataSurvey2019}       

\begin{figure}
\centering
\includegraphics[width=\columnwidth]{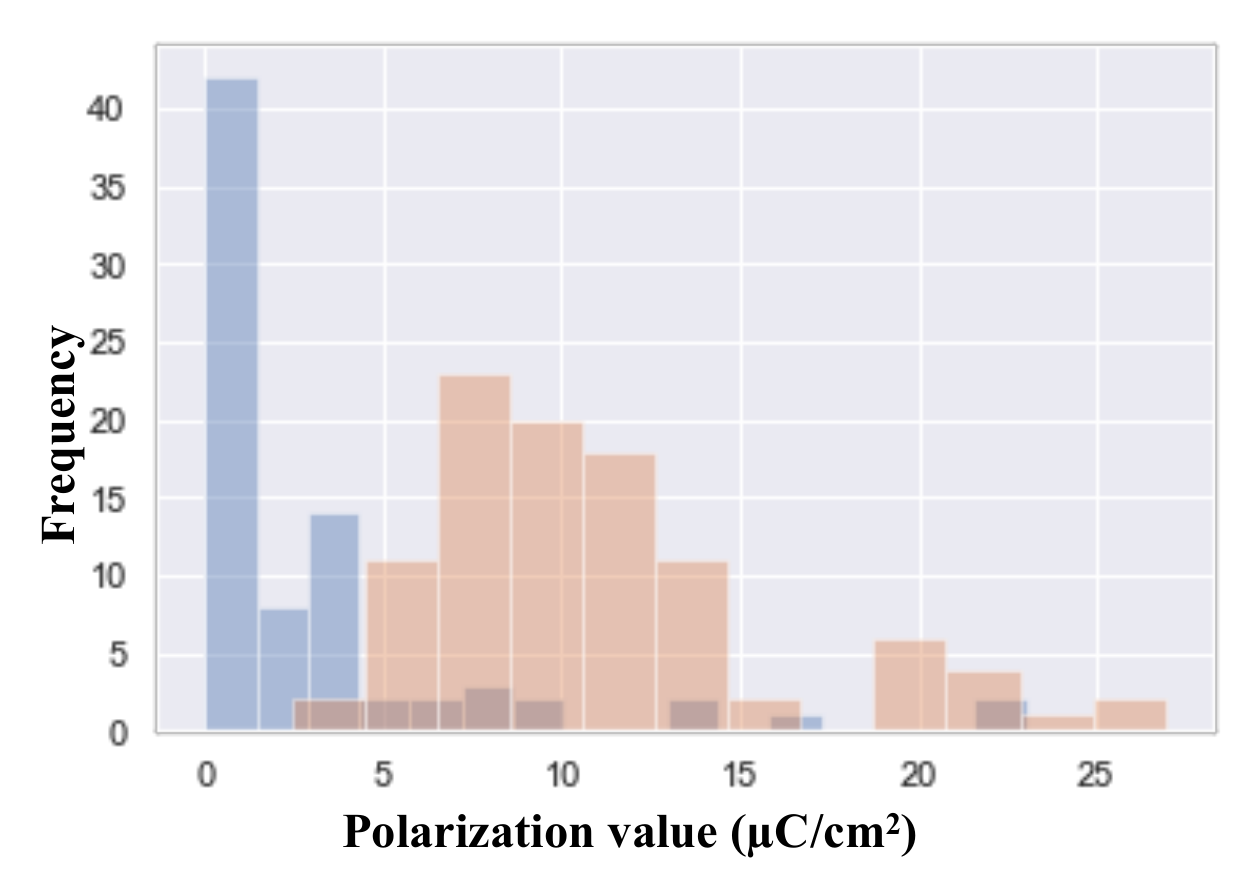}
\caption{
Histogram partitioning of data points in Dataset II, including both the original (blue bars) and newly generated entries
with importance sampling (orange bars), by the magnitude of their spontaneous polarization.}
\label{fig:fig3_org_ferro}
\end{figure}

Since the list of already known ferroelectric compounds has been greatly exhausted in the process of assembling Dataset I,
we employed an \emph{importance sampling}\cite{neal2001annealed} based approach to generate additional synthetic data points 
with high polarization values --- which was accomplished by assigning greater importance to high-polarization instances that 
have low occurrences in Dataset I. 
Such statistical strategies, allowing for an infusion of additional samples around low occurrence training data instances, 
are widely used to address the problem of unbalanced training data in different ML modeling strategies.\cite{chawla2002smote}
To generate a balanced version of Dataset I using the importance sampling approach, we had to define a range of polarization 
magnitudes that is considered important for our analysis. 
We set the lower boundary of this range to 5 $\mu$C/cm$^2$, as Fig.~\ref{fig:fig2_org_ferro} 
clearly shows that there are few data instances present in Dataset I with polarizations greater than this value.
This lower boundary to generate the samples can be picked in a randomized manner. 
However, if significant number of original data points do not satisfy the criteria, the new samples generated will be biased towards low frequency values. 
Lack of variance in such newly generated data points will not help to improve the accuracy of the constructed ML model.
Hence, a use of higher threshold of polarization value, e.g., 0 $\mu$C/cm$^2$ or above (only less than 10 data points 
available in Dataset I with such polarization values) will not yield meaningful samples falling under the range of polarization values of interest.
High dimensionality of the involved descriptor space ($\sim$1,000) introduces difficulties in generating coherent samples from such 
large multivariate distributions.
We note that this large space was initially constructed by combining all possible descriptors and then filtered down
on the ML model building stage to contain only descriptors that were highly ranked by their importance.
However, during the sample generation stage, we kept all descriptors in their original form to avoid missing any relevant information.
To reduce the dimensionality of the descriptor space, we applied Principal Component Analysis (PCA)\cite{abdi2010principal} 
that projects high-dimensional features (descriptors) to orthogonal, mutually independent dimensions called 
principal components (PC) by performing an eigen-decomposition of the covariance matrix in the descriptor space. 
The resulting eigenvectors correspond to the orthogonal PC dimensions. 
An advantage of using a linear transformation model like PCA is that an inverse transformation can be performed to move from the 
mutually independent PC space to the high-dimensional descriptor space.
After transforming the high-dimensional descriptor space to low-dimensional PC space with PCA, 
we generated multiple random samples for the independent PC dimensions. 
On that sampling step, we first used a kernel density estimate (KDE)\cite{kdebook2018} model (with gaussian kernels) 
to approximate the individual distributions of the independent PCs.
We then produced 100 new samples from this KDE distribution and projected them back to the original high-dimensional descriptor space,
utilizing the inverse transformation property of PCA. 
This resulted in 100 new data points that are likely to have polarization values above 5 $\mu$C/cm$^2$.
Orange colored bars in Fig.~\ref{fig:fig3_org_ferro} represent the distribution of polarization magnitudes corresponding to these new 100 synthetically created data instances.
These new samples were combined with the (unbalanced) Dataset I to create a relatively balanced dataset, called in what follows `Dataset II,' that includes 
more data instances possessing high polarization values.
Note that fictitious samples generated by this procedure were used exclusively to train the ML models,
while any test sets consisted only of unaltered original data collected for the actual materials.

\begin{figure*}
\includegraphics[width=\textwidth]{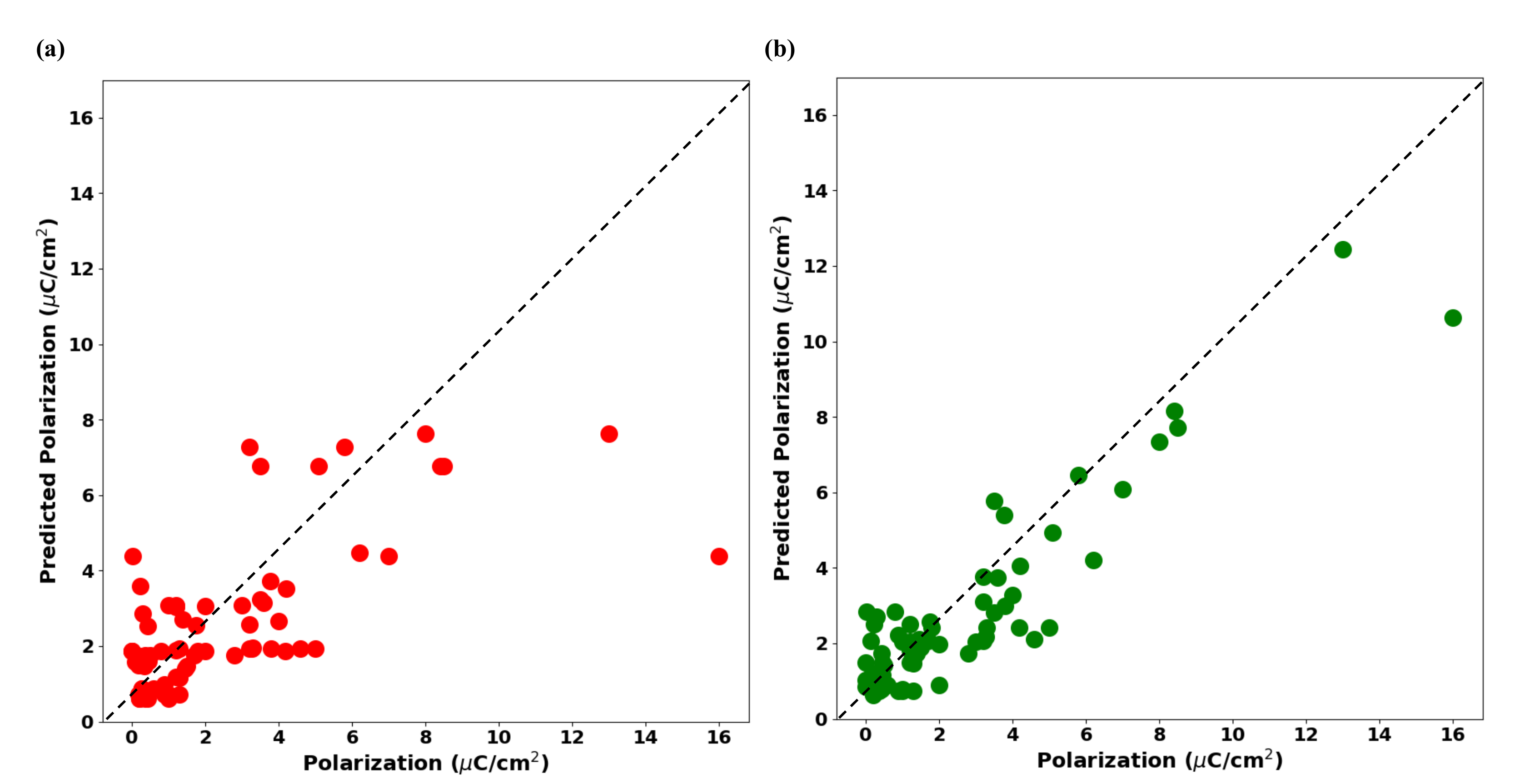}
\caption{Predicted vs.\ actual polarization values, with the former produced by models constructed on (a) Dataset I and (b) Dataset II.
No synthetically created samples in Dataset II (see Sec.~\ref{MLM_imp}) were used to compute predicted polarization values.
}
\label{fig:parity_plot}
\end{figure*}

\section{Results and Discussion}\label{para:org_Results}

\subsection{Predicting Ferroelectric Polarization}
The first iteration of ML models for predicting ferroelectric polarization was built on Dataset I, with one model constructed utilizing molecular-level descriptors 
only and the other using only crystal-level descriptors.
The average RMSEs of these ML models for predicting spontaneous polarization were close to 2.16 $\mu$C/cm$^2$.
On the next iteration, an improved model utilizing 10 most important descriptors taken from the combined set of molecular- and crystal-level descriptors
was constructed, with the average RMSE of 1.96 $\mu$C/cm$^2$.
The range of polarizations was 1--15 $\mu$C/cm$^2$, with most falling below 5 $\mu$C/cm$^2$.
To improve the predictive accuracy of the original ML models, 
we constructed the third iteration of models employing Dataset II.
It is evident that utilization of the importance sampling approach to un-bias Dataset I 
improved the predictive quality of the models, reducing the average test set RMSE error to 1.29 $\mu$C/cm$^2$,
while also decreasing the overfitting.
The parity plots for the best models constructed using Datasets I and II are shown in Fig.~\ref{fig:parity_plot}.
We note that it is possible to create different variants of Dataset II by changing the lower boundary of the 
`important polarization' range to a value that is higher than 5 $\mu$C/cm$^2$.
However, due to much smaller number of reported structures with polarization $>10$ $\mu$C/cm$^2$, 
samples generated for datasets with large `important polarization' cut-offs  will be too similar in nature, 
therefore not contributing any substantial new information to the dataset.
Consequently, any ML models constructed using such datasets will not have their predictive accuracy significantly improved.
A different strategy may involve building ML models on datasets that aggregate only those compounds that have 
the same mechanism for developing ferroelectricity.
That approach can further improve prediction accuracy, but at present such partitioning reduces the original dataset to 
subsets of less than 20 entries in size.

\begingroup
\begin{table*}
\tabcolsep=0.2cm
\renewcommand{\arraystretch}{1.2}
\caption{Potential organic ferroelectrics for which only 2 structural phases are present in the CSD. 
Molecular formulas, canonical SMILEs and predicted spontaneous polarization values (in $\mu$C/cm$^2$) are provided.
The polarization values have been reported up to three digits after the decimal to avoid rounding errors.}
\resizebox{\textwidth}{!}
{\begin{tabular}{lccc}
\hline
                                             & Molecular Formula & Canonical SMILE & $|\mathbf{P}_s|$  \\     
\hline
& C$_{10}$H$_{15}$N$_{5}$O & CC(CCNC1=NC=NC2=C1NC=N2)CO & 1.476 \\
& C$_{10}$H$_{6}$O$_{8}$ & C1=C(C(=CC(=C1C(=O)O)C(=O)O)C(=O)O)C(=O)O & 2.069 \\
& C$_{13}$H$_{21}$NO$_8$ & CCOC1C=C(OC(C1NC(=O)C)C(C(CO)O)O)C(=O)O & 1.747 \\
& C$_{13}$H$_{21}$N$_5$ & C1CCC(C1)C2=CC(=NC(=N2)N)N3CCC(C3)N & 7.098 \\
& C$_{13}$H$_{7}$N$_3$O & C1=CC(=CN=C1)OC2=CC(=C(C=C2)C\#N)C\#N & 1.586 \\
& C$_{15}$H$_{10}$N$_2$O$_6$ & COC(=O)C1=C2C(=NC3=C(N2)C(=O)C=C(C3=CO)O)C(=O)C=C1 & 1.651 \\
& C$_{16}$H$_{14}$F$_3$NO$_2$ & C1CC2=CC3=C(C4=C2N(C1)CCC4)OC(=O)C=C3C(F)(F)F & 2.024 \\
& C$_{17}$H$_{19}$N$_5$O$_4$ & C1C(C(OC1N2C=NC3=C2N=C(N=C3OCC4=CC=CC=C4)N)CO)O & 1.701 \\
& C$_{18}$H$_{20}$N$_6$O$_3$ & CC1(N=C(N=C(N1C2=CC(=CC=C2)OCC3=CC(=CC=C3)[N+](=O)[O-])N)N)C & 1.572 \\
& C$_{19}$H$_{11}$NO$_3$ & C1=CC=C(C=C1)C=CC2=NOC3=C2C(=O)C4=CC=CC=C4C3=O & 1.750 \\
& C$_{19}$H$_{12}$BF$_{2}$N$_{3}$O & B(C1=C(N=CC=C1)F)(C2=C(N=CC=C2)F)OC3=CC=CC4=C3N=CC=C4 & 1.987 \\
& C$_{19}$H$_{22}$N$_{2}$O$_3$C$_2$H$_6$O & CC1=NOC(=C1)CCCCCCCOC2=CC=C(C=C2)C3=NC(CO3)CO & 1.493 \\
& C$_{19}$H$_{30}$O$_7$ & CC(C)(CO)CO.C1=CC=C2C(=C1)C(=O)OC2=O.C(CCCO)CCO & 1.403 \\
& C$_2$H$_3$NO$_3$ & C(=O)(C(=O)O)N & 4.347 \\
& C$_2$H$_5$NO & CC(=O)N & 3.342 \\
& C$_{20}$H$_{22}$O$_7$H$_2$O & CC(=O)CC1CCC(=O)C=CC(=O)OC(CCC(=O)C=CC(=O)O1)CC(=O)C & 1.704 \\
& C$_{20}$H$_{27}$F$_3$O$_3$ & C(CCC(F)(F)F)CC=CCC=CCC=CC=CC(=O)CCCC(=O)O & 1.671 \\
& C$_{21}$H$_{36}$O$_{5}$ & CCCCCC(C=CC1C(CC(C1CC=CCCCC(=O)OC)O)O)O & 1.238 \\
& C$_{22}$H$_{12}$N$_2$O$_4$ & C1=CC=C2C(=C1)C(=O)OC(=N2)C3=CC=C(C=C3)C4=NC5=CC=CC=C5C(=O)O4 & 1.851  \\
& C$_{23}$H$_{25}$F$_3$N$_2$O$_3$ & C1CN(CC(C1C2=CC=C(C=C2)O)F)C3CCN(C3=O)CC4=CC=C(C=C4)OC(F)F & 1.783 \\
& C$_{24}$H$_{28}$N$_6$O$_2$ & CCCCOC1=CN=CC(=C1)C2=NN=C3N2C4=C(C=CC(=C4)CN5CCOCC5)N=C3C & 1.298 \\
& C$_{24}$H$_{34}$O$_{15}$ & CC(=O)OCC1C(C(C(C(O1)C2(C(CCOC2COC(=O)C)OC(=O)C)O)OC(=O)C)OC(=O)C)OC(=O)C & 1.805 \\
& C$_{24}$H$_{35}$N$_3$ & CC1=C2CCCCN2C(=C1)C3=CC=C(C=C3)N(C)CCCN4CCCCC4 & 7.098 \\
& C$_{25}$H$_{31}$N$_3$O$_2$ & CC(C)N1CCC(CC1)NC(=O)C2=CC3=CC=CC=C3N2CC4=CC(=CC=C4)OC & 1.298 \\
& C$_{26}$H$_{22}$BNO$_3$ & [B-]12([N+]3=C(C=CC=C3C4=C(O1)C=CC(=C4)C)C5=C(O2)C=CC(=C5)C)C6=CC=C(C=C6)OC & 1.213 \\
& C$_{27}$H$_{16}$ & C1C2=C3C(=CC=C2)C4=CC=CC5=C4C(=CC=C5)C3=C6C1=C7C=CC=CC7=C6 & 6.804  \\
& C$_{28}$H$_{20}$O$_{6}$ & C1=CC(=CC=C1C=CC2=C3C(=CC(=C2)O)OC(=C3C4=CC(=CC(=C4)O)O)C5=CC=C(C=C5)O)O & 1.876 \\
& C$_{28}$H$_{28}$B$_2$N$_4$ & [B-]12(CC([B-](C3=[N+]1C=CN3C)([N+]4=C2N(C=C4)C)C5=CC=CC=C5)C6=CC=CC=C6)C7=CC=CC=C7 & 6.639 \\
& C$_{29}$H$_{45}$NO$_5$ & CC(C)C(=O)OC(C)C=CC(=O)NC1CCC(CC1)CC=C(C)C=CC2CC3(CC(O2)(C)C)CO3 & 1.255 \\
& C$_{32}$H$_{24}$O$_3$ & CC1=C2C3=C(C=CC=C3O)C(C4=C2C(=CC=C4)O1)CC(=C5C6=CC=CC=C6C7=C5C=C(C=C7)O)C & 1.298 \\
& C$_{32}$H$_{26}$O$_3$ & COC1=CC=C(C=C1)C2=CC(=C(C=C2)C3=CC=CC=C3)C4=CC=CC=C4OC5=CC=C(C=C5)OC & 1.214 \\
& C$_{32}$H$_{36}$N$_2$O$_5$ & CC1CC=CC2C3C(O3)(C(C4C2(C(=O)C=CC(=O)C(C(=C1)C)O)C(=O)NC4CC5=CNC6=CC=CC=C65)C)C & 1.262 \\
& C$_{36}$H$_{35}$N$_3$ & C1C2CC3CC1CC(C2)C4=C3C=C5C=CC6=C7C=CC8=CC9=C(C1CC2CC(C1)CC9C2)N=C8C7=NC6=C5N4 & 6.787 \\
& C$_{36}$H$_{54}$O$_2$ & CC(=CCCC(=CCCC(=CCOC1=CC=C(C=C1)OCC=C(C)CCC=C(C)CCC=C(C)C)C)C)C & 1.252 \\
& C$_{39}$H$_{54}$ & CC1=C(C(CCC1)(C)C)C=CC(=CC=CC=CC=CC=C(C)C=CC=C(C)C=CC2=C(CCCC2(C)C)C)C & 7.058 \\
& C$_{40}$H$_{34}$N$_{2}$O$_6$ & CNCCC1=CC2=CC=CC=C2N(C3=CC=CC=C31)C.C1=CC=C2C(=C1)C=C(C(=C2C3=C(C(=CC4=CC=CC=C43)C(=O)O)O)O)C(=O)O & 1.376 \\
& C$_{41}$H$_{31}$N & CC1=CC=C(C=C1)C2=CC(=CC3=CC=CC=C32)N(C4=CC=C(C=C4)C5=CC=CC=C5)C6=CC=C(C=C6)C7=CC=CC=C7 & 6.795 \\
& C$_{41}$H$_{74}$O$_2$ & CCCCCCCCCCCCCCCC(=O)OCCC(C)CCC=C(C)CCC=C(C)CCC=C(C)CCC=C(C)C & 1.515 \\
& C$_{44}$H$_{56}$O$_{10}$ & CC1CC2C3CCC4=CC(=O)C=CC4(C3C(=O)CC2(C1(C(=O)CO)O)C)C.CC1CC2C3CCC4=CC(=O)C=CC4(C3C(=O)CC2(C1(C(=O)CO)O)C)C & 1.446 \\
& C$_{45}$H$_{78}$O$_2$ & CCCCCCCCC=CCCCCCCCC(=O)OC1CCC2(C3CCC4(C(C3CC=C2C1)CCC4C(C)CCCC(C)C)C)C & 1.346 \\
& C$_{46}$H$_{24}$ & C1=CC=C2C(=C1)C3=CC=CC=C3C4=CC5=C(C=C24)C6=CC7=CC8=C(C=C7C=C65)C9=C8C=C1C2=CC=CC=C2C2=CC=CC=C2C1=C9 & 6.744 \\
& C$_{48}$H$_{38}$ & CC(C)(C)C1=CC2=C3C4=CC=CC=C4C5=CC=CC=C5C3=C6C=C(C=C7C6=C2C(=C1)C8=C7C9=CC=CC=C9C1=CC=CC=C18)C(C)(C)C & 6.853 \\
& C$_{48}$H$_{42}$O$_2$ & CC(C)(C)C(C\#CC1=CC=CC=C1C\#CC2=CC=CC=C2)(C3=CC=C(C=C3)C(C\#CC4=CC=CC=C4C\#CC5=CC=CC=C5)(C(C)(C)C)O)O & 1.569 \\
& C$_6$H$_{11}$N$_5$O$_4$ & C(C(CNC(=O)O)CN=[N+]=[N-])NC(=O)O & 1.715 \\
& C$_7$H$_{14}$O$_7$ & C(C(C(C(C(C(=O)CO)O)O)O)O)O & 2.134  \\
\hline
\end{tabular}}
\label{table:shortlisted}
\end{table*}
\endgroup

\subsection{Shortlisting potential organic ferroelectrics}
The developed ML models for predicting the magnitude of spontaneous polarization in molecular- and polymer-based compounds
were integrated into a strategy for shortlisting potential candidate materials (within the same general chemical space) 
that may exhibit ferroelectricity.
The guiding principles outlined in Sec.~\ref{intro} were used to identify the most attractive targets from the list
of 1,034,175 organic compounds present in the CSD.\cite{groom2016cambridge}
In order to be shortlisted as a potential ferroelectric, a compound must have at least two different structural phases 
that belong to centrosymmetric and polar space groups, respectively.
The presence of both centrosymmetric and polar polymorphs within the phase space of same material is a condition
that is similar to the well-known cases of ceramic-oxide ferroelectrics, such as BaTiO$_3$ and PbTiO$_3$, where spontaneous polarization emerges 
as a (centrosymmetry breaking) structural distortion of the non-polar high-temperature phase upon cooling into a low-temperature polar phase.
On the other hand, this condition cannot be considered as sufficient or even necessary for the development of ferroelectricity.
\begin{figure}
\centering
\includegraphics[width=\columnwidth]{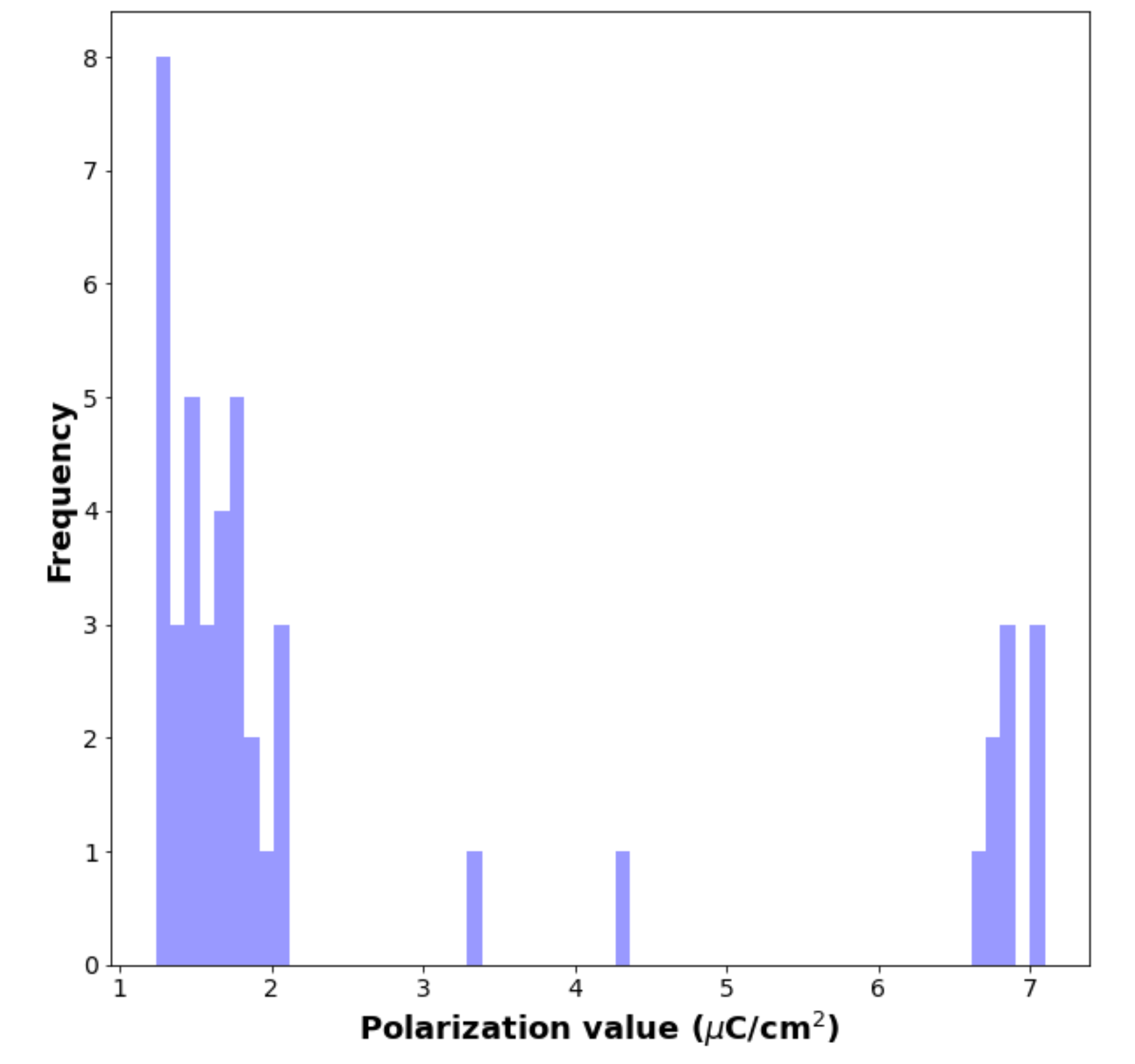}
\caption{
Histogram showing predicted polarizations of compounds from Table~\ref{table:shortlisted}.
%Polarizations for DIPA-Cl\cite{louis2018polarization} and 100\% crystalline $\beta$-PVDF\cite{Nakhmanson2004,Nakhmanson2005PRB} are shown for comparison.
%Note that typical polarizations of (partially amorphous) poled experimental samples of $\beta$-PVDF or PVDF/TrFE are at least twice as weak \cite{Nakhmanson2005PRB}.
}
\label{fig:fig5_org_ferro}
\end{figure}
Utilizing this simple criterion, we identified a total of 8,997 compounds, which is a very large number that needs to be reduced further.
In order to accomplish that, we utilized the following constraints:
\begin{enumerate}
\item{Compounds containing heavy elements, such as Mo, As, Au, Ba, Dy, Eu, Ga, Gd, Ge, Hg, La, Nd, Pb, Rh, Ru, Sb, Se, U,Ir, Os, Pd, Pt, Re, Ta, Sm, Tb, Te, U, W, Yb, Y, Sb, Tc, Tm, Co, Nb, In, V and Be, were discarded;}
\item{Compounds containing toxic elements, such as Cd, Pb, Th, Hg and Sb, were discarded;}
\item{Hydrates and charged complexes were discarded;}
\item{Only compounds containing elements belonging to the first row of the periodic table were retained, 
although this step may be skipped in future investigations to include more diverse compounds into the mix.}
\end{enumerate}
After the application of these filters, a total of 4,932 compounds remained in the final list. 
For these compounds, the spontaneous polarization could be quickly predicted by utilizing the previously developed ML models,
thus further reducing the number of attractive targets for emerging ferroelectricity. 
On further downsampling steps, atomistic-level calculations, such as those employing density-functional theory (DFT),
may be required --- e.g., to compute the relative energy differences between the phases and evaluate their potential closeness,
as indicated in criterion (\emph{ii}) of Sec.~\ref{intro}.
\begin{figure*}
\centering
\includegraphics[width=\textwidth]{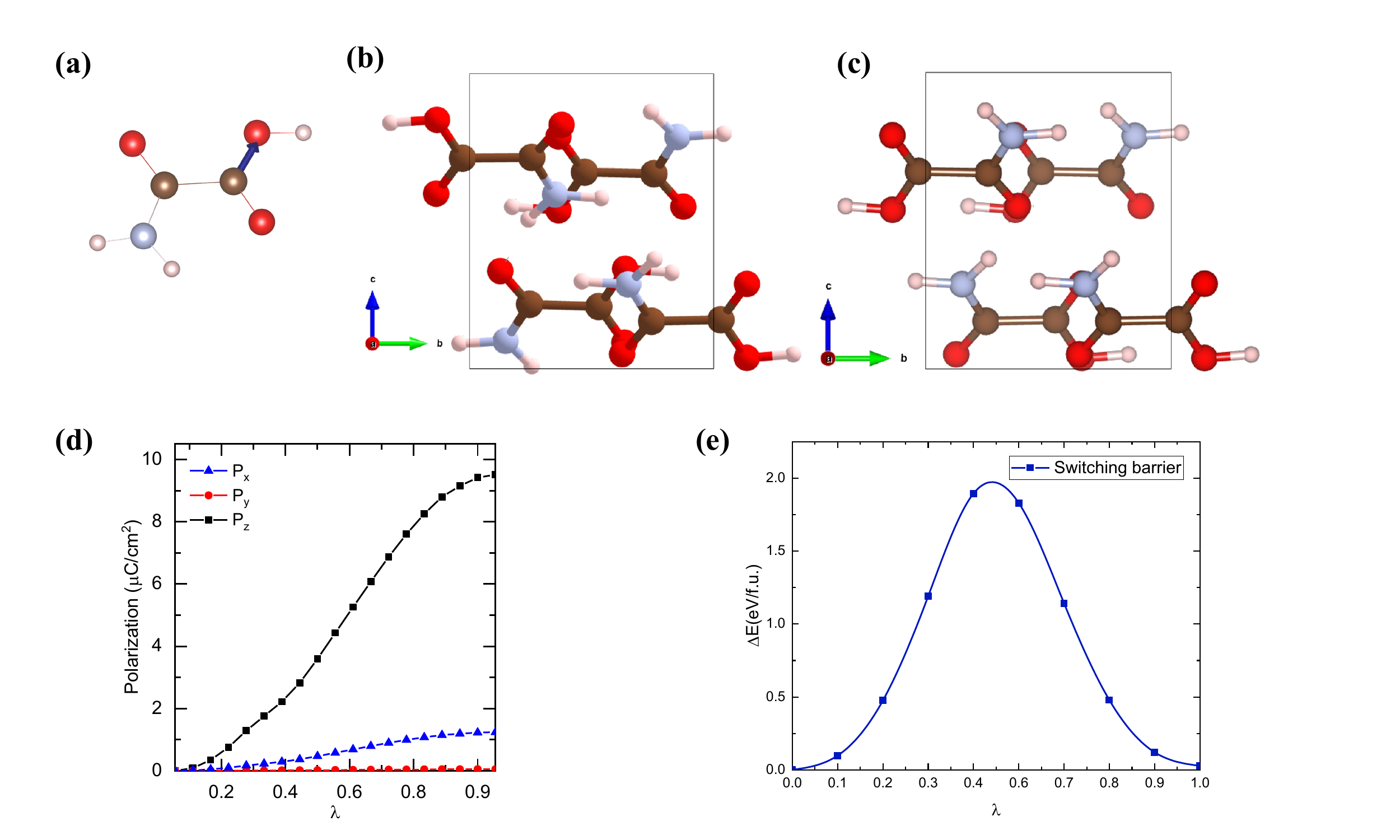}
\caption{(a) C$_2$H$_3$NO$_3$ molecular unit schematics, (b) bulk non-polar phase, (c) polar phase of the oxamic acid, respectively.
(d) Polarization components P$_{x}$, P$_{y}$, and P$_{z}$ along crystallographic axes $a$, $b$ and $c$, are plotted respectively with (e) corresponding energy differences as functions of the polar distortion parameter $\lambda$, computed using DFT.
See text for more details.
}
\label{fig:fig6_org_ferro}
\end{figure*}
However, prediction of the system polarization requires structural descriptions (i.e., atomic coordinates and unit cell parameters) for each of the compound phases,
which are not available as free-access features of CSD.
Instead, to demonstrate the predictive capabilities of the developed ML models in evaluating the system spontaneous polarization, 
we have created a short list of compounds that exhibit exactly two structural phases --- one centrosymmetric and the other one polar --- with the same lattice type. 
This list contains 45 individual compounds that are collected in Table~\ref{table:shortlisted}. 
All of these compounds can also be found in CSD, but (naturally) they are not present in Dataset I.
Space group assignments for the centrosymmetric and polar polymorphs for each of the compounds are provided in the Supplementary Material 
and can be easily analyzed further, e.g., for the presence of group-subgroup relationships between the two.
For this list, we were able to generate the canonical SMILEs and molecular formulas for all the compounds
using PubChem\cite{kim2019pubchem} and OpenBabel\cite{o2008pybel,o2011open} libraries.
In turn, that allowed us to estimate the magnitudes of their spontaneous polarizations, which are also included in Table~\ref{table:shortlisted}.
The polarization magnitudes vary approximately between 1.238 and 7.098 $\mu$C/cm$^2$.
These values are obtained by averaging over 300 predictions by the model constructed using Dataset II to avoid any discrepancy 
(changes in magnitudes when the model is run multiple times) arising due to limited dataset size.
A histogram plot of all predicted polarizations is shown in Fig.~\ref{fig:fig5_org_ferro}.
As identified in the Table and Figure, there are nine compounds with predicted polarization of $\sim$7 $\mu$C/cm$^2$,
i.e., close in magnitude to experimental values reported for partially crystalline $\beta$-PVDF.\cite{Nakhmanson2005PRB,Meng2019}
It is noteworthy that the quality of these predictions can be improved further if the compound structures 
are obtained directly from CSD, rather than created using PubChem and OpenBabel.
Again, at this point, the list entries could be evaluated further on an individual basis, according to criteria (\emph{ii}) and (\emph{iii}) of Sec.~\ref{intro},
to develop more accurate predictions of their propensity to exhibit ferroelectricity.
However, since such activities involve performing DFT-based simulations and thus would be rather
computationally expensive, they fall beyond the scope of this investigation.
We point out that such computations (including more accurate estimation of spontaneous polarization exhibited by the polar phases and 
other related properties) are standard for DFT-based computational packages and could be easily carried out when necessary or desired.

\subsection{A validation example}

In this section we include an example showcasing how such an investigation, utilizing DFT to estimate the polar properties, can be conducted.
The DFT\cite{hohenberg1964density} calculations were performed using the Vienna \textit{ab initio} simulation package (VASP)\cite{kresse1996efficient} within the projector-augmented-wave (PAW) scheme.\cite{blochl1994projector} 
The Perdew-Burke-Ernzerh (PBE) functional was used to treat exchange and correlation interactions\cite{perdew1996generalized}. 
The cutoff energy was set at 520.0 eV and a $k$-point mesh of 4$\times$6$\times$6 was used for the computations. 
All the structures were fully relaxed until all forces on all atoms are smaller than 0.001 eV/\AA. 
Berry phase formalism was used to calculate spontaneous polarization $\mathbf{P}_s$.\cite{resta1992theory}
Here, the system of interest is oxamic acid (C$_2$H$_3$NO$_3$), that was predicted to have the polarization of 4.347 $\mu$C/cm$^2$ by the ML model (see Table~\ref{table:shortlisted}).
A sketch of the (planar) C$_2$H$_3$NO$_3$ molecular unit is presented in Fig.~\ref{fig:fig6_org_ferro}(a).
The direction of the unit net dipole moment is expected to be along the edge C atom bonded with an (-OH) functional group.
This is because O is more electronegative than N present in the C-NH$_2$, located at the other edge of the molecular unit.
The bulk polar structure\cite{dugarte2019structure} is arranged of four molecular units, as shown in Fig.~\ref{fig:fig6_org_ferro}(b), and belongs to the $Cc$ space group.
To estimate the spontaneous polarization for this system, the corresponding reference non-polar centrosymmetric structure was constructed.\cite{capillas2011new}
The non-polar structure belongs to the $P2_{1}/c$ space group and is connected to the polar $Cc$ structure by a 180$^{\circ}$ rotation of 
two of the four molecular units around the axis including the C--C bond, as shown in Fig.~\ref{fig:fig6_org_ferro}(c).
A polar distortion parameter $\lambda$ is introduced to represent the transition path from the non-polar ($\lambda$ = 0) to polar structure ($\lambda$ = 1). 
%
%We note that the intermediate configurations are only designed (may not be physical) to exclude the polarization quanta as often miscounted for computing polarization using the Berry phase formalism.
%
The $x$, $y$, and $z$ components of the spontaneous polarization as a function of $\lambda$ are plotted in Fig.~\ref{fig:fig6_org_ferro}(d).
The components, P$_{x}$, P$_{y}$, and P$_{z}$ are along crystallographic axes $a$, $b$ and $c$, respectively.
Although the difference in the ground-state energies for polar and non-polar phases are very close, the energy differences as obtained by performing a Nudge Elastic Band (NEB) computation suggests that the switching barrier (Fig.~\ref{fig:fig6_org_ferro}(e)) is high for the intermediate structures, $\sim$ 2 eV per molecular unit.
Polarization components along $x$ (1.240 $\mu$C/cm$^2$) and $y$ ($\sim$0 $\mu$C/cm$^2$) are minimal due to the canted arrangements of the molecular unit
dipole moments, while $P_{z}$ is 9.510 $\mu$C/cm$^2$.
Although the DFT estimate for the system polarization is larger than the ML model prediction, it nonetheless falls remarkably close by the order of magnitude.
A possible source of discrepancy may be in an overly simplified (cubic) generic structure of the polar phase created from the SMILE representation by the ML model,
which is different from the actual structure shown in Fig.~\ref{fig:fig6_org_ferro}(b).
Finally, we note that the polarization estimate for the oxamic acid molecular crystal ($\sim$10 $\mu$C/cm$^2$) is a relatively large value
for a polymer- or molecular-based ferroelectric, e.g., in comparison to typical polarizations exhibited by the PVDF family of materials,
as well as some other compounds.\cite{Nakhmanson2005PRB,louis2018polarization,ghosh2018first}

\section{Conclusions}
In conclusion, we have compiled datasets containing reports on molecular and
polymer-based ferroelectric compounds from the available literature. 
Through various data analytics techniques, we have identified several descriptors that are critical for understanding polar properties of such systems and developed families of ML models that could predict the magnitude of spontaneous polarization in them.
We have also applied data mining to shortlist potential organic ferroelectrics based on their
polymorphism, symmetry and chemical makeup, and predicted the values of spontaneous polarization for them.
Further investigations including checking the stability of the ground-state configurations and height of the energy barriers 
between structural phases can be easily accomplished and will significantly enhance the presented analysis, 
but are beyond the scope of this project as it is currently envisioned. 
Overall, the developed general prescription, employing both data-driven sampling techniques and ML models 
describing polar properties of organic crystals (but applicable to inorganic systems as well), 
helps us improve our understanding of mechanisms governing 
the emergence of such complex material functionalities, as well as identify potential candidates possessing
such functionalities for further in-depth investigations.

\begin{acknowledgements}
A.G. acknowledges the hospitality of Los Alamos National Laboratory, where this project was initialized. 
This work was supported by the U.S.\ DOE NNSA under Contract No.\ 89233218CNA000001 through the LANL Institute for Materials Science (IMS), 
and in part through the Center for Integrated Nanotechnologies, a U.S.\ DOE Office of Basic Energy Sciences user facility in partnership with the 
LANL Institutional Computing Program for computational resources. 
A.G.\ also acknowledges her current funding support from the U.S.\ DOE, Office of Science, Office of Basic Energy Sciences Data, 
Artificial Intelligence and Machine Learning at DOE Scientific User Facilities.
\end{acknowledgements}
%

%%%REFERENCES%%%
\bibliography{references} %You need to replace "rsc" on this line with the name of your .bib file
\bibliographystyle{rsc} %the RSC's .bst file

\subsection{Contributions}
A.G.\ performed all computations and analysis and prepared all the figures. 
A.G., S.N.\ and J.-X.Z.\ developed the project idea. 
S.H., D.T., E.S.\ and H.T.\ participated in curation of experimental reports and code development. 
A.G.\ and S.N.\ wrote the paper with inputs from everyone. \\

\subsection{Conflicts of Interest}
There are no conflicts of interest to declare.

\subsection{Additional Information}
The author(s) declare no competing interests.

\end{document}